\documentclass[12pt]{article}
\usepackage{graphicx}
\usepackage{url}
\usepackage{cite}
\usepackage[margin=1.25 in]{geometry}
\usepackage[colorlinks = true, linkcolor = blue, urlcolor = blue,
      citecolor = blue, anchorcolor = blue]{hyperref}

\newcommand\pubnumber{SLAC-PUB-17516}
\newcommand\pubdate{March 2020}

\def\SLAC{SLAC,
    Stanford University, Menlo Park, California 94025 USA}
\def\doeack{\footnote{Work supported by the US Department of Energy,
                     contract DE--AC02--76SF00515.}}

\def\Title#1{\begin{center} {\Large #1 } \end{center}}
\def\Author#1{\begin{center}{ \sc #1} \end{center}}

\newcommand\pubblock{\rightline{\begin{tabular}{l} \pubnumber\\
         \pubdate \end{tabular}}}
\newenvironment{Abstract}{\begin{quotation} \begin{center}
                       ABSTRACT
     \end{center}\bigskip  }{\end{quotation}}
\newenvironment{Presented}{\begin{quotation} \begin{center} 
             CONTRIBUTED TO\end{center}\bigskip 
      \begin{center}\begin{large}}{\end{large}\end{center} \end{quotation}}

\def\Acknowledgements{\bigskip  \bigskip \begin{center} \begin{large}
             \bf ACKNOWLEDGEMENTS \end{large}\end{center}}

\def\beq{\begin{equation}}
\def\eeq#1{\label{#1}\end{equation}}
\def\eeqn{\end{equation}}

\newenvironment{Eqnarray}%
   {\arraycolsep 0.14em\begin{eqnarray}}{\end{eqnarray}}
\def\beqa{\begin{Eqnarray}}
\def\eeqa#1{\label{#1}\end{Eqnarray}}
\def\eeqan{\end{Eqnarray}}
\def\CR{\nonumber \\ }

\def\leqn#1{(\ref{#1})}

\let\bar=\overbar

\def\etal{{\it et al.}}

\def\lsim{\mathrel{\raise.3ex\hbox{$<$\kern-.75em\lower1ex\hbox{$\sim$}}}}
\def\gsim{\mathrel{\raise.3ex\hbox{$>$\kern-.75em\lower1ex\hbox{$\sim$}}}}

\def\O{{\cal O}}

\def\del{\partial}
\def\Dslash{\not{\hbox{\kern-4pt $D$}}}
\def\dslash{\not{\hbox{\kern-2pt $\del$}}}

\def\Dlr{\mathrel{\raise1.5ex\hbox{$\leftrightarrow$\kern-1em\lower1.5ex\hbox{$D$}}}}

\def\ee{e^+e^-}
\def\sstw{\sin^2\theta_w}

\def\msb{{\bar{\scriptsize M \kern -1pt S}}}

\def\drb{{\bar{\scriptsize D \kern -1pt R}}}

\makeatletter
\def\section{\@startsection{section}{0}{\z@}{5.5ex plus .5ex minus
 1.5ex}{2.3ex plus .2ex}{\large\bf}}
\def\subsection{\@startsection{subsection}{1}{\z@}{3.5ex plus .5ex minus
 1.5ex}{1.3ex plus .2ex}{\normalsize\bf}}
\def\subsubsection{\@startsection{subsubsection}{2}{\z@}{-3.5ex plus
-1ex minus  -.2ex}{2.3ex plus .2ex}{\normalsize\sl}}

\renewcommand{\@makecaption}[2]{%
   \vskip 10pt
   \setbox\@tempboxa\hbox{\small #1: #2}
   \ifdim \wd\@tempboxa >\hsize     
       \small #1: #2\par          
     \else                        
       \hbox to\hsize{\hfil\box\@tempboxa\hfil}
   \fi}

\makeatother

\begin{document}
\begin{titlepage}
\pubblock

\vfill
\Title{Precision Theory of Electroweak Interactions}
\vfill
\Author{ Michael E. Peskin\doeack}
 \medskip
\begin{center} 

  \SLAC 
\end{center}
\vfill
\begin{Abstract}
As a part of the celebration of 50 years of the Standard Model of
particle physics, I present a brief history of the precision theory of
electroweak interactions.   I emphasize in particular the theoretical
preparations for the LEP program and the prediction of $m_t$ and $m_h$
from the electroweak precision data.
\end{Abstract} 
\vfill
\begin{Presented}
  The Standard Model at 50 Years\\
  Case Western Reserve University,  June 1-4, 2018
  \end{Presented}
\vfill
\end{titlepage}

\hbox to \hsize{\null}


\tableofcontents

\def\thefootnote{\fnsymbol{footnote}}
\newpage
\setcounter{page}{1}

\setcounter{footnote}{0}

\section{Introduction}

The Standard Model of the weak interaction went through three distinct
stages of development.   The first was the era of confusion,
speculation, and, eventually, insight that ended with the 1967-8 papers
of Weinberg and Salam~\cite{Weinberg,Salam}.   The second, initiated
by
the discovery of 't
Hooft and Veltman that non-Abelian gauge theories are
renormalizable~\cite{tHVrenorm}, was an experimental program that
narrowed down the gauge group of the model to $SU(2)\times U(1)$ and
verified the predictions of the model at the 10\% level.    The stage
ended with the measurement of parity violation in deep inelastic
electron scattering at SLAC~\cite{DES}.  At that point, the evidence
for the Standard Model had become sufficiently compelling to motivate
the award of the 1979 Nobel Prize to Sheldon Glashow, Abdus Salam, and
Steven Weinberg.
It is worth remembering that this prize was given before the actual discovery of
the $W$ and $Z$ bosons in 1981 by the UA1 experiment~\cite{UAone}.

Still, another stage was needed. The Standard Model had not yet been
stress-tested with measurements of high precision.  The accuracy of the measurements
was not sufficient even  to convince some that the weak interactions
were based on an exact, not an approximate, gauge
symmetry~\cite{Bjorken:1980gb,Abbott:1981re,Fritzsch:1981zh,Chen:1982yj}.  Beyond
this, once the community began to consider the Standard Model as
established, high-precision tests of this model would become a method
to  search for additional fundamental physics interactions  hidden
at very short distances. 

In the 1990's, the LEP and SLC experiments realized this promise,
verifying the predictions of the Standard Model at the level of
fractions of a percent.    Alain Blondel describes the
experimental achievements in his contribution to this
volume~\cite{Blondel}.
This program also required an unprecedented theoretical effort to
produce calculations of the predictions of the Standard Model at a
level matched to the experimental precision.   The purpose of this
contribution is to review that theoretical achievement, the creation
of a precision electroweak theory.

It is just an accident that I was asked to give the presentation on
this topic at the symposium.
Bryan Lynn, an important contributor to this field, was scheduled
for this talk but could not attend
the symposium.   I have never done a high-precision
electroweak calculation.  But I hope that I can represent this history
appropriately in the discussion to follow.

\section{Earliest steps}

The history of precision electroweak calculation actually begins
before the concept of the electroweak theory and, to some extent, even
before the idea of the $W$ boson.  In 1957, Toichiro Kinoshita and
Alberto Sirlin were inspired by the new $V$--$A$ theory of
charge-changing weak interactions~\cite{FG,MS} and worked to make the
predictions of that theory more precise. In advance of the postulate
of $V$--$A$, Louis Michel had put forward a formalism for testing the
form of the weak interaction using measurements on muon
decay~\cite{Michel}.  In a series of papers, Kinoshita and Sirlin 
computed the radiative corrections to the Michel formula to first
order in $\alpha$~\cite{Kinoshita:1957zz, Kinoshita:1958ru}.
Their results exposed many theoretically interesting
features that would appear again in further precision quantum field
theory calculations.   The calculations contain infrared divergences
that cancel in a way that is reminiscent of the cancellations in pure
QED.  But they also contain a  logarithmic ultraviolet divergence due to
the non-renormalizable nature of the 4-fermion interaction.  This term
needed to be absorbed into a new phenomenological parameter.  This ultraviolet
problem could not be resolved without a deeper underlying theory.

Two decades later, after the invention and the first evidence for the
$SU(2)\times U(1)$ electroweak theory, it became possible to move this
program forward.  Martinus Veltman and Alberto Sirlin picked up again
the dream of turning the theory of weak interactions into one
buttressed by  high
precision calculation.

\begin{figure}
\begin{center}
\includegraphics[width=0.40\hsize]{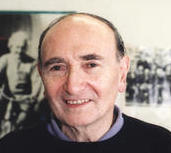}\ \
\includegraphics[width=0.43\hsize]{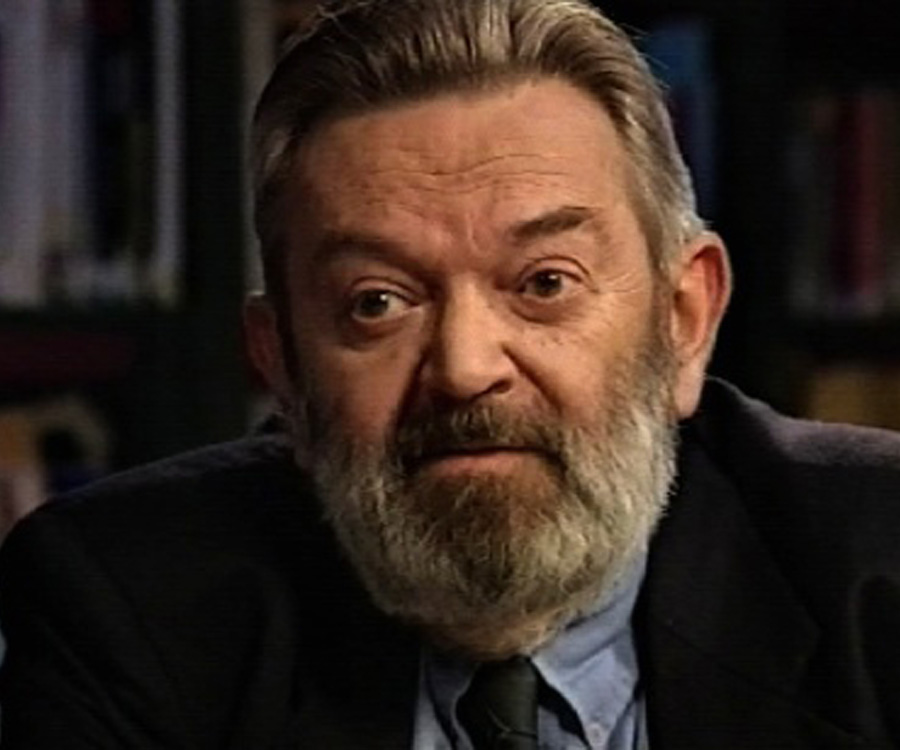}\ \
\end{center}
\caption{Leaders in the precision calculation of the predictions of
  the electroweak theory:  (left)  Alberto Sirlin; (right) Martinus Veltman.}
\label{fig:SirlinVeltman}
\end{figure}

Veltman and Giampiero Passarino took the first major step in
this direction.   In 1978, they computed the complete set of 1-loop radiative
corrections to the cross section for $\ee\to\mu^+\mu^-$ in the
electroweak theory~\cite{Passarino:1978jh}.  In the process, they
provided methods that made more general 1-loop electroweak
calculations feasible.  In particular, Passarino and Veltman presented
a method to reduce all of the integrals that appear in this
calculation to a small set of ``master integrals''.  Once these
integrals have been evaluated, the work of computing radiative
corrections is reduced to pure --- although very complex --- algebraic
manipulation.   The general formulae for the 3- and 4-point master
integrals are not straightforward to obtain.   But, fortunately,
Veltman could ask his former graduate student  Gerard  't~Hooft to take time
off from theoretical theory  to help in finding the appropriate
tricks. The resulting
paper, which completely solved the problem of the evaluation of
1-loop 
Feynman integrals with massive denominators, is also foundational in
this enterprise~\cite{tHooft:1978jhc}.

In the 2000’s, the Passarino-Veltman method was  enhanced by
additional tricks due to Bern, Dixon, Dunbar, and Kosower~\cite{Bern:1994zx} 
and Ossola, Pittau, and Papadopoulos~\cite{Ossola:2006us}.
These methods allow general reductions of diagrams
for processes with arbitrarily many external particles.
The combination of methods is amenable to automation, and
this is now achieved in codes
such as \verb+MadGraph5_aMC@NLO+~\cite{Alwall:2014hca, Degrande:2018neu}.
Today, even experimenters can generate predictions at 1-loop accuracy!

\section{What, precisely, is $\sstw$?}
\label{sec:sstw}

Once one knows how to compute the diagrams, there is another
important conceptual problem to be solved.  This is the question of
how to organize the renormalization of the $SU(2)\times U(1)$ model.

Sirlin also had dreams of completing the 1-loop analysis of weak
interaction processes.   His focus was much broader than high energy
$\ee$ reactions; it included the computation of radiative corrections
for neutrino reactions  and other low-energy
probes, and also the idea, new in the 1970's, of building precise
tests of grand unified theories.    In \cite{Marciano:1980be} and
\cite{Sirlin:1981yz},
Sirlin and William Marciano began a series of papers that addressed
these questions.   Their results called attention to many of the
important physical effects of electroweak radiative corrections. In particular,
the running of the QED coupling
constant $\alpha$ has a surprising large influence, giving a 3\%
correction to the tree-level prediction for the $W$ boson mass and an order of
magnitude shift in the prediction for  the grand unification scale.

Computing radiative corrections for such a wide variety of processes
requires a systematic approach to renormalization.   At the tree
level, the electoweak theory has three parameters on which all of its
predictions depend---the $SU(2)\times U(1)$ coupling constants $g$ and
$g'$ and the Higgs field vacuum expectation value $v$~\cite{Higgs}.   The
electroweak theory is
renormalizable.  This means that, at the 1-loop level, these three
parameters receive divergent corrections, but, once these three parameters are
adjusted back to finite values, the expressions for  all other observables of the theory
are also rendered finite.   However, these finite corrections depend
on the definitions chosen for $g$, $g'$, and $v$. There are many, many
possibilities to define these quantities.  Unless one makes a definite
choice, the values of the radiative corrections are ambiguous and
untestable.

Marciano and Sirlin introduced the procedure of defining the
parameters of the electroweak theory by an ``on-shell''
renormalization method.  They defined the underlying parameters of
the theory by giving a special role to the quantities, 
\beq
     \alpha(m_Z^2) \ , \quad   G_F \ , \quad \sin^2\theta_w|_{M-S} \equiv 1 -
     m_W^2/m_Z^2  \ , 
     \eeq{MSchoice}
   the running value of the QED coupling at the $Z$ mass scale, the
   Fermi constant defined from muon decay, and the ratio of the $W$
   and $Z$ boson masses.  These quantities are not precisely
   observables, but they are very closely connected to quantities
   measured in weak interaction experiments.  Marciano and Sirlin defined the
   underlying parameters of the
   electroweak theory so that these quantities 
 should receive zero radiative corrections after renormalization.
 This definition supplies a set of counterterms that can then be used to
 cancel the divergences in all other predictions of the theory.  A
 straightfoward way to implement this is to write 1-loop expressions for
 electroweak observables in terms of the parameters in
 \leqn{MSchoice}.   Then, renormalizability implies that these
 expressions must be finite.

The parameters $\alpha(m_Z^2)$ and $G_F$ were already accurately known
before the start of the precision electroweak measurements at the $Z$
pole.   The best choice for the third parameter is less clear.
Two other definitions of $\sstw$ proved to be useful in interpreting
the results of the $Z$ measurements.   These make use of two
quantities that would be measured especially accuately in those
experiments, the value $m_Z$  of the $Z$ boson mass  and  the value
$A_\ell$ of the polarization asymmetry of the $Z$-lepton couplings.
Specifically,  
\beqa
  \sin^2\theta_0 \quad & \mbox{defined\ by} &\quad  \sin^22\theta_0 \equiv
  {\alpha(m_Z^2)\over \sqrt{2} G_F m_Z^2} \CR
  \sin^2\theta_* \quad & \mbox{defined\ by} & A_\ell \equiv
  \quad {  1/4 - \sin^2\theta_*
   \over   1/4 - \sin^2\theta_* +
   2 \sin^4\theta_*} \ .
 \eeqa{twomoresstw}

 All three definitions of
 $\sstw$ coincide at the tree level.  Then, by the considerations of
 the previous paragraph, the difference between any two of these values of
 $\sstw$ generated by 1-loop corrections  is a finite quantity that is a prediction of the electroweak
 theory.   In fact,  $\sin^2\theta_0$ and  $\sin^2\theta_*$ differ only at the
 part-per-mil level, and the use of either give values for the $Z$
 branching ratios and asymmetries at the tree
 level that are already within 1\% of the  complete expressions.
The quantity $\sin^2\theta_*$ was generalized by Dallas Kennedy and
Bryan Lynn to a gauge-invariant running
$\sin^2\theta_*(Q^2)$~\cite{Kennedy:1988sn}, which similarly
accurately encodes
the electroweak corrections to 
the differential cross sections for $\ee\to f\bar f$ processes.

 More recently, the Particle Data Group has chosen to quote
 the value of $\sstw$ using  $\msb$ subtraction, with the parameters set
 by the best fit to the global corpus of electroweak data~\cite{PDG}.

 \section{The $Z$ lineshape}
 \label{sec:lineshape}

 The LEP and SLC experiments were designed to sit on the $Z$ resonance
 and measure $\ee\to Z$ production and individual $Z$ decays,
 hopefully with the highest statistics possible.   This brought up
 another important question that required an improved conceptual
 understanding:
What is the precise form of the $Z$ resonance line-shape as a function
of the $\ee$ center of mass energy?   In particular, where is the peak
of the $\ee\to Z\to f\bar f$ cross section relative to the position
$m_Z$ of
the $Z$ boson pole, and how does the peak cross section compare to
that in the simplest approximation?

At leading order, the $Z$ is described as a 
Breit-Wigner resonance with width $\Gamma_Z$ computed from the
$Z$-fermion couplings.   However, this does not give a line-shape even
close to the observed one.
The line-shape is distorted by the effect of initial state
radiation of photons from the incoming electrons and positrons,
shifting the position of the peak and producing a long tail  of the
resonance toward higher values of the CM energy.  The usual figure of
merit for the magnitude of initial-state radiation is
\beq
   \beta = {2\alpha\over \pi} (\log {s\over m_e^2} - 1)   = 0.108
   \quad\mbox{at} \  s = m_Z^2 \ .
   \eeq{betaval}
  However, since the $Z$ is narrow, the relative distortion of the
  resonance  is larger,
  of order
  \beq
  - \beta \ \log{m_Z\over \Gamma_Z} = - 40\% \ .
  \eeqn
On the other hand, in order to test the electroweak prediction for the $Z$
width,  the line-shape of the resonance needs
  to be known to part-per-mil accuracy.

  This could be done using an approach introduced in 1987 by Victor
  Fadin and Eduard Kuraev.   They argued that multiple photon
  emissions could be accounted by viewing the hard electrons and
  photons as partons of the electron and using the formalism for  parton
  evolution in QCD  (in this context, the Gribov-Lipatov
  equation~\cite{Gribov}) to sum over real and virtual emissions
  systematically~\cite{Fadin,Fadin2}.   Since the electron is an elementary
  particle, one could also obtain the initial condition for the parton
  evolution equation from perturbation theory and thus  have a
  complete solution.   The $Z$ line-shape could then be obtained from
  an overlap of these parton distributions, convolved with a hard
  matrix element that included higher-order diagrams with virtual $W$
  and $Z$ bosons.  Using  the Fadin-Kuraev solution as a starting point,
  later analyses added higher-order photon resummation and the
  complete finite order $\alpha^2$ contributions~\cite{NandT,Bardinline}.
  
\section{The Yellow Book}

With all of the conceptual elements in place, it was still necessary
to carry out the hard work of turning these ideas into precise
theoretical predictions.  Too many people contributed to this effort
to list them all in this short review, but I would like to call
attention to the major ``schools'' that contributed, those of Manfred
B\"ohm in Wurzburg (whose students include Wolfgang Hollik and Ansgard
Denner)~\cite{Bohm1,Bohm2},   Frits Berends in Leiden (whose students include
Ronald Kleiss and Wim Beenakker)~\cite{Berends1,Berends2}  and Dmitri Bardin in Dubna (whose
students include Tord Riemann)~\cite{Bardin1,Akhundov}.   Over the
past 15 years, these calculations have been extended to full 2-loop order
in the electroweak interactions by Ayres Freitas, Tord Riemann, and their
collaborators; see \cite{Dubovyk1,Dubovyk2} and references therein.

\begin{figure}
\begin{center}
\includegraphics[width=0.50\hsize]{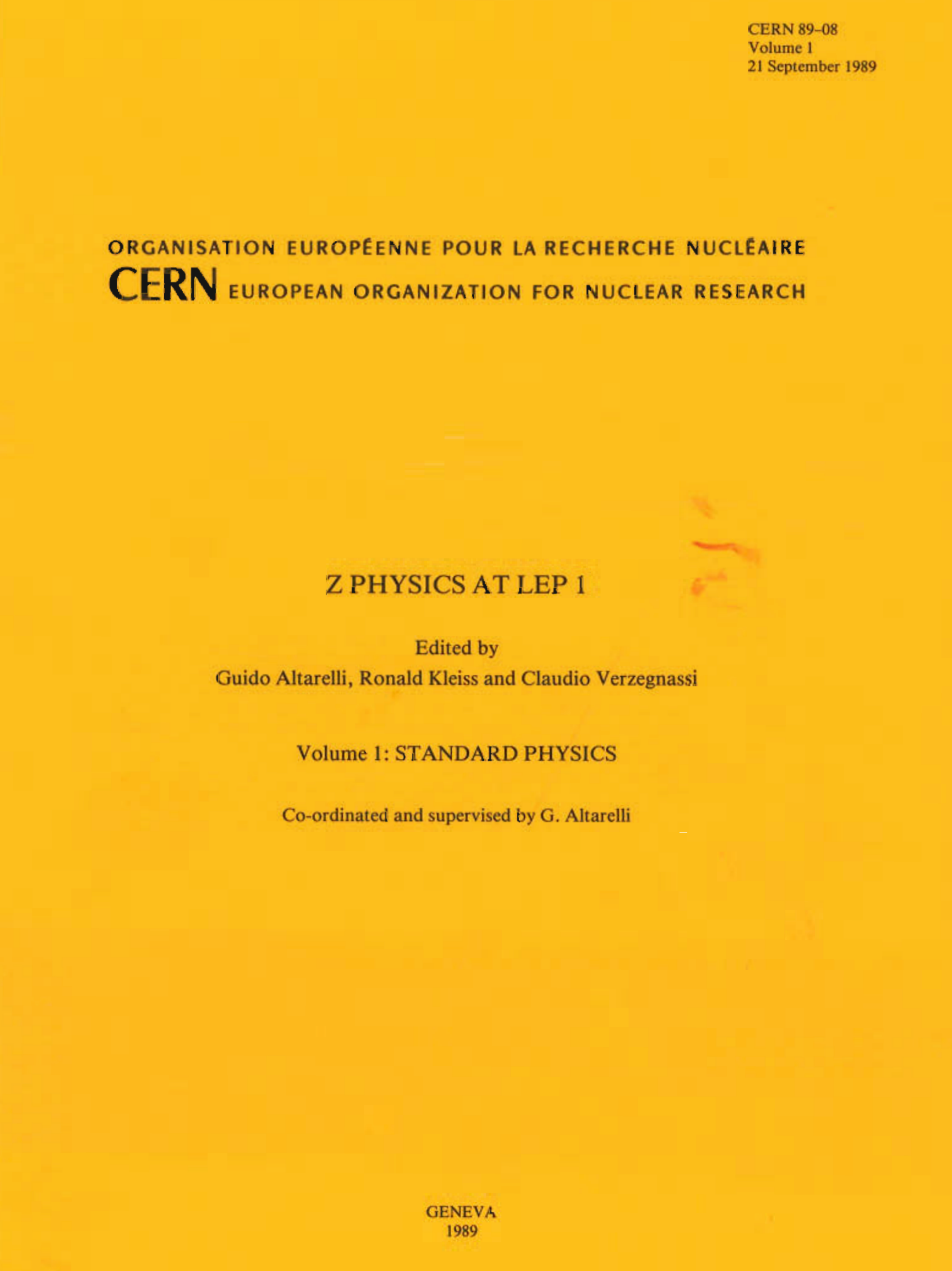}
\end{center}
\caption{The Yellow Book,  CERN-YELLOW-89-08.}
\label{fig:AltarelliYellowBook}
\end{figure}

A milestone in the progress of the precision theory was the LEP Yellow
Book ``Z Physics at LEP 1'', edited by Guido Altarelli, Ronald Kleiss,
and Claudio Verzegnassi~\cite{Yellow}.   Altarelli marshalled the
efforts of the whole European theory community to ensure that all
aspects of the $Z$ resonance physics would be worked out to 1-loop
precision and the results explained in detail.

\begin{figure}
\begin{center}
\includegraphics[width=0.99\hsize]{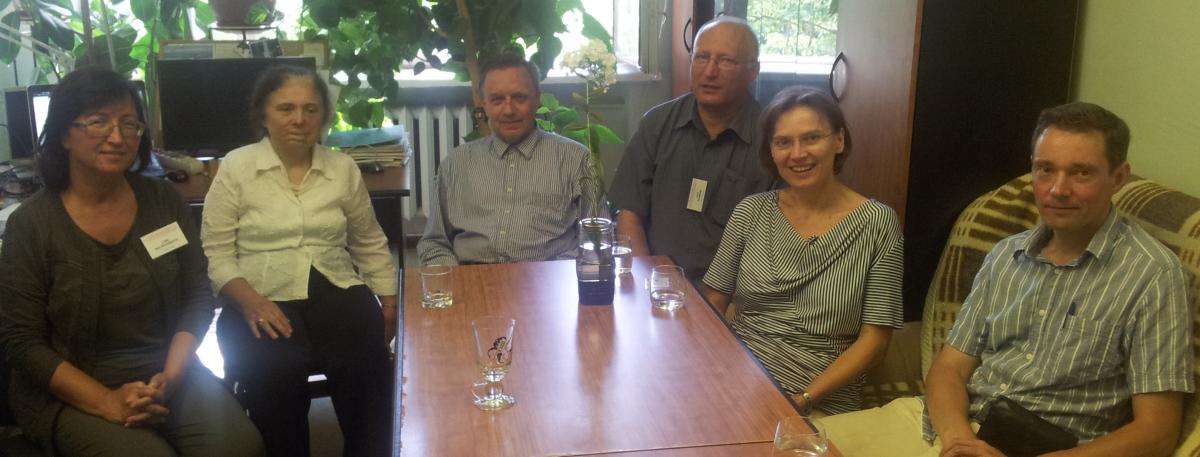}
\end{center}
\caption{The team of ZFITTER authors: left to right, Lida
  Kalinovskaya, Pena Christova, Dmitri Bardin, Tord Riemann, Sabine
  Riemann, Andrej Arbuzov.}
\label{fig:ZFITTER}
\end{figure}

As the experiments began to take data, these  theory
predictions needed to be presented as  event generators whose output
could be directly compared to the observed distributions.   The two
most influential of these were ZFITTER, developed by a group at Dubna
and DESY-Zeuthen led by Bardin~\cite{ZFITTER1,ZFITTER2},
and KORALZ, developed by Stanislaw
Jadach, Bennie Ward, and Zbigniew Was~\cite{KORALZ}.
Figure~\ref{fig:ZFITTER} shows the ZFITTER team in a relaxed moment.

\section{Precision electroweak measurements at the $Z$}

\begin{figure}
\begin{center}
\includegraphics[width=0.70\hsize]{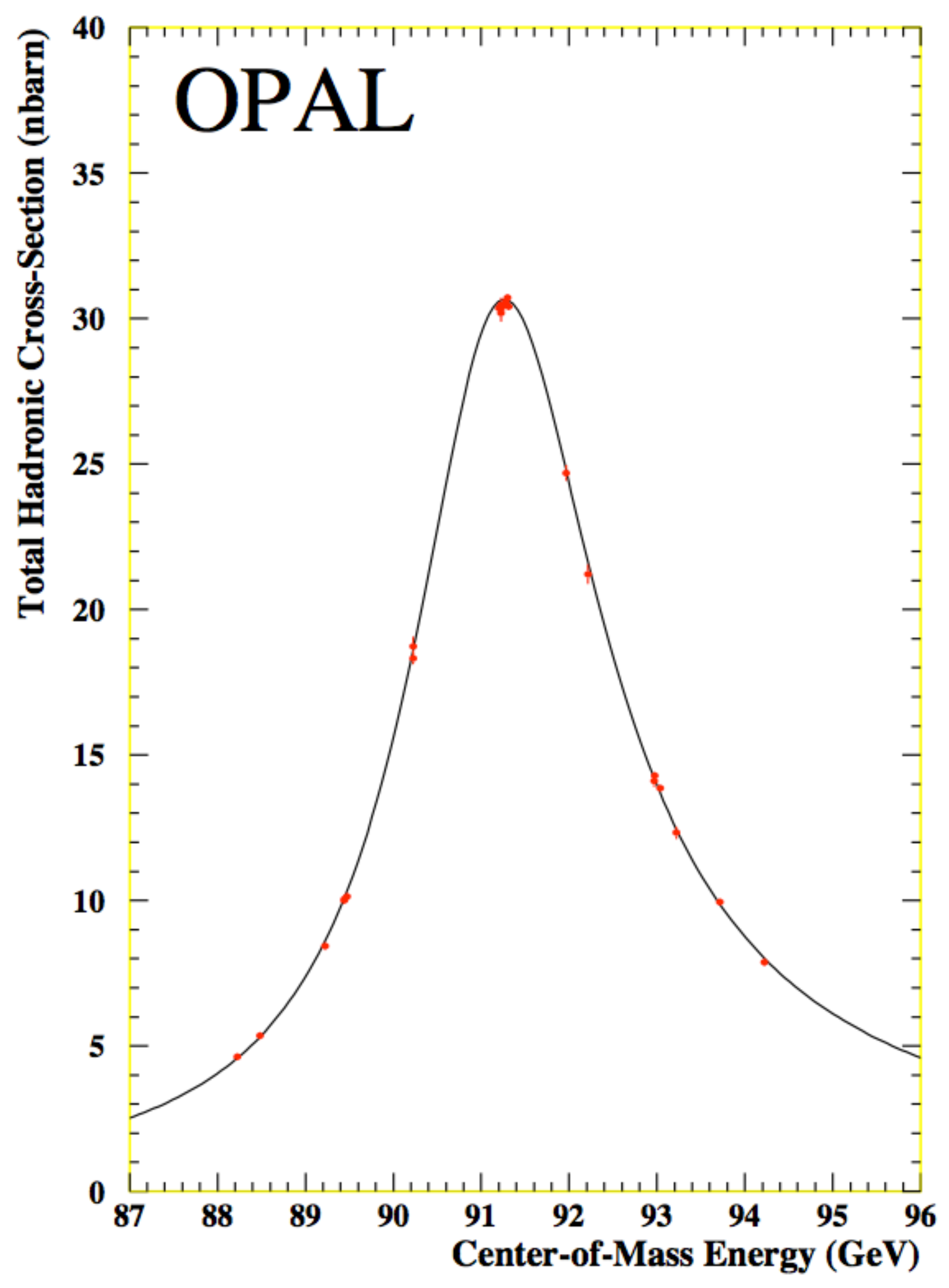}
\end{center}
\caption{Measurements of the $Z$ boson lineshape, and comparison to
  theory; figure courtesy of T. Mori based on \cite{OPALZ}.}
\label{fig:ZshapeOPAL}
\end{figure}

\begin{figure}
\begin{center}
\includegraphics[width=0.80\hsize]{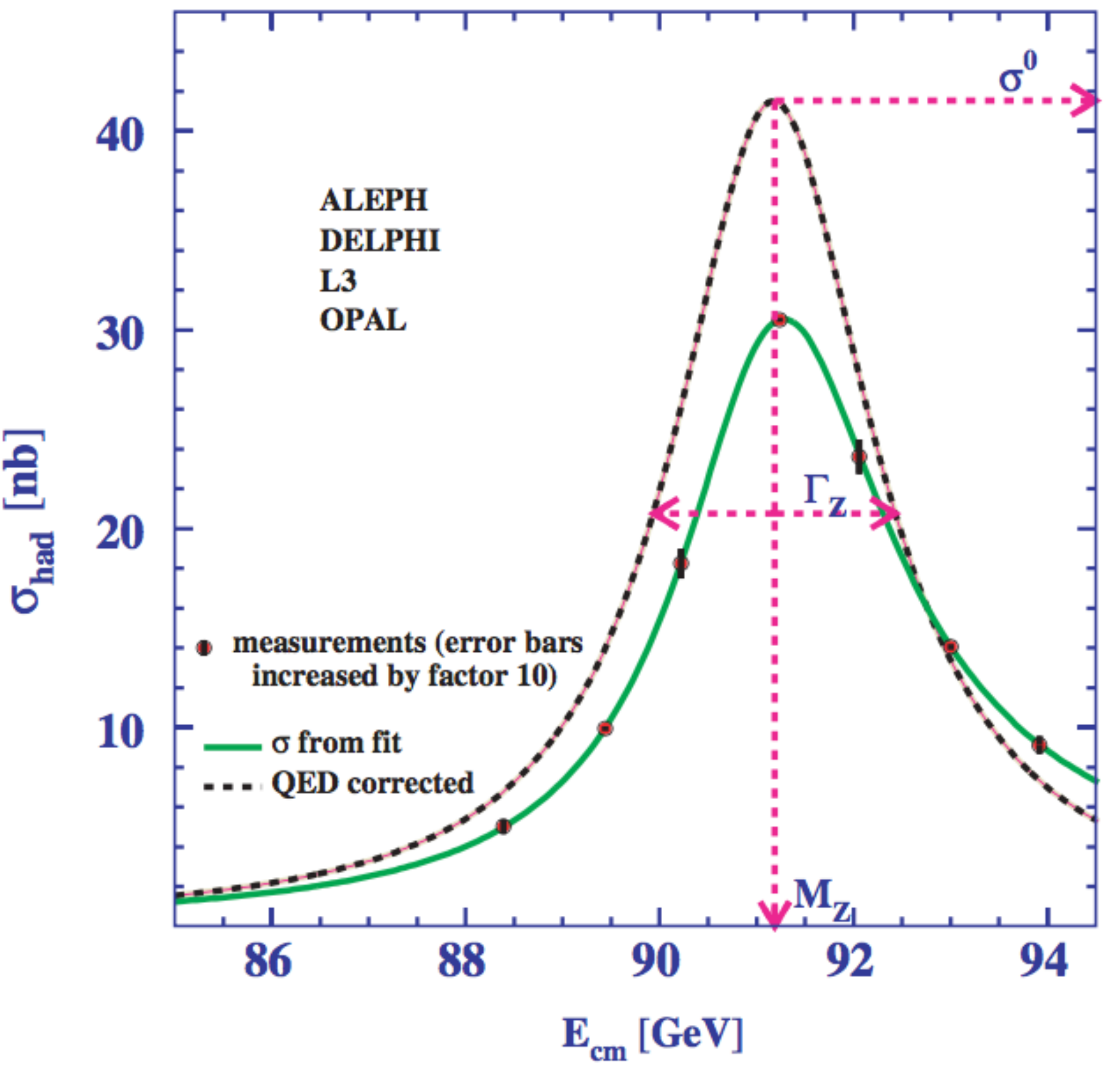}
\end{center}
\caption{Composite figure representing measurements of the $Z$ line
  shape by the four LEP experiments ALEPH, DELPHI, L3,
  OPAL~\cite{LEPEWZ}.  The experimental errors have been increased by a
  factor 10 to make them visible.  The dotted curve shows the ideal
  resonance shape before inclusion of initial-state photon radiation.}
\label{fig:ZshapeLEP}
\end{figure}

Alain Blondel's article in this volume describes the experiments
from viewpoint of one of the participants~\cite{Blondel}.  However, it seems
appropriate here to highlight some of the most impressive comparisons
between theory and experiment.   In general, in this section, when I
quote 
experimental values of observables, these are taken 
from the 2005 summary paper of the LEP Electroweak
Working Group~\cite{LEPEWZ}.   When I quote theoretical predictions,
these are taken from the
Standard Model best-fit values given in the article of Erler and
Freitas in the {\it Review of Particle Physics}~\cite{PDG}.

 In Section~\ref{sec:lineshape}, I emphasized the subtlety of
 predicting the $Z$ resonance line shape.   The result of the LEP
 measurements of the lineshape are shown in Figs.~\ref{fig:ZshapeOPAL}
 and \ref{fig:ZshapeLEP}.    Figure~\ref{fig:ZshapeOPAL} shows the
 point-by-point hadronic cross sections measured by the
 OPAL experiment, compared to the high-order theory~\cite{OPALZ}.
 Figure~\ref{fig:ZshapeLEP}, which represents the composite
 measurements by the four LEP experiments ALEPH, DELPHI, L3, and OPAL as
 a 7-point scan~\cite{LEPEWZ}, also shows more clearly the effect of initial-state
 radiation. The line is noticeably shifted to higher energies, to
 account for the energy of the photons radiated before the $\ee$
 annihilation.  More notably, the peak height of the resonance is
 decreased by 30\%.

 The width of the underlying Breit-Wigner resonance is affected by QCD
 and electroweak corrections and also, possibly, by the decay of the
 $Z$ to new light particles.  The  width extracted from the analysis
 of LEP data  is  $2.4955 \pm 0.0023$~MeV.  (This value reflects very
 recent updates concerning the measurements of the LEP luminosity and
 the 2-loop calculation of Bhabha scattering to
 which these measurements are compared~\cite{Janotlumi,Janot}.)    The
 Standard Model prediction is 2.4965~MeV.

\begin{figure}
\begin{center}
\includegraphics[width=0.50\hsize]{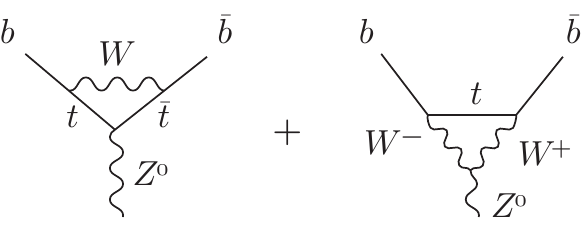}
\end{center}
\caption{Top quark diagrams contributing to $\Gamma(Z\to b\bar b)$. }
\label{fig:bbar}
\end{figure}

 A particularly interesting component of the $Z$ width is the decay to
 bottom quarks.   The $b_L$  is the $SU(2)$ partner of the $t_L$, so
 in 
 models in which the top quark interacts with new particles outside
 the Standard Model, the $b_L$ will usually also feel some of these
 effects.  Even within the Standard Model, there is a significant
 radiative correction, due to the diagrams in Fig.~\ref{fig:bbar},
 that gives a $-2\%$ correction to the $Z\to b\bar b$ partial
 width~\cite{Akhundov}.  The observable 
 \beq
 R_b = \Gamma(Z\to b\bar b)/\Gamma(Z\to \mbox{hadrons}) \ .
 \eeq{Rbdef}
offers a chance to observe these effects.
 From the lowest-order couplings without radiative corrections, one
 would expect $R_b = 0.220$.    The full Standard Model prediction is
 $0.21562$.   The composite value from  the four LEP experiments
 is $R_b = 0.21629 \pm 0.00066$. This is in excellent agreement with the
 Standard Model and puts a strong constraint on models of new 
interactions specific to the top quark. 

As I have noted in Section~\ref{sec:sstw},  the
 polarization asymmetries of the quark and lepton couplings to the $Z$
 are affected only slightly by radiative corrections.   However, the
 overall pattern of these asymmetries is a striking qualitative
 prediction of the Standard Model.  The polarization asymmetries are
 defined as
 \beq
    A_f  =   {\Gamma(Z\to f_L\bar f_R) - \Gamma(Z\to f_R\bar f_L)
          \over   \Gamma(Z\to f_L\bar f_R) + \Gamma(Z\to f_R\bar f_L)}
            \ .
            \eeq{Af}
   Within the Standard Model, the asymmetries depend strongly on the
   electroweak quantum numbers,
   \beq
      A_f =  \quad  0.15\ \mbox{for\ $\ell$}\ , \quad 0.63\ \mbox{for\
        $u$}\ , \quad
      0.94\ \mbox{for\ $d$} \ .
      \eeq{Afvalues}

    The value of the leptonic asymmetry $A_\ell$, which is relatively
    modest, can be measured in several different ways. First, it can
    be measured from the forward-backward asymmetry in $\ee\to f\bar
    f$ on the $Z$ resonance.  At the tree level, this is related to $A_f$ in a relatively
    simple way,
    \beq
    A_{FB}(\ee\to f\bar f) =  {3\over 4}  A_e A_f \ .
    \eeq{AFBrelation}
    One should note that the very small predicted  value of this asymmetry in
    $\ee\to \mu^+\mu^-$ (1.7\%) is increased by about a factor of 2
    due to radiative corrections.    In the process $\ee\to
    \tau^+\tau^-$, the final-state polarization of the $\tau$ can be
    measured from the weak-interaction asymmetries in the $\tau$
    decays.   At central values of the production angle $\cos\theta$, the observed
    polarization reflects $A_\tau$; however, at forward and backward
    values of $\cos \theta$, the observed asymmetry reflects the
    forward-backward asymmetry induced by $A_e$.  More generally, the
    $\tau$ polarization is given at tree level by~\cite{JW}
    \beq
         P_{\tau^-}(\cos\theta) = -  {A_\tau (1 + \cos^2\theta) + {3\over 4}
           A_e \cos\theta \over  (1 + \cos^2\theta) + {8\over 3}  A_{FB}^\tau
           \cos\theta  }  \ ,
         \eeq{Ptaubyangle}
         or, since $A_{FB}^\tau$ is small,
    \beq
         P_{\tau^-}(\cos\theta) \approx  -  \biggl( A_\tau + 2 A_e
           {\cos\theta\over  (1 + \cos^2\theta)} \biggr) \ .
     \eeq{Ptaubyangleapprox}
   Thus, this observable separately measures $A_e$ and $A_\tau$.   In
   the LEP measurements, the two values turned out to be compatible,
   as expected from Standard Model lepton universality,
   \beqa
            A_\tau &=&  0.1439 \pm 0.0043  \CR
            A_e &=&  0.1498 \pm 0.0049  \ .
            \eeqa{Ataue}
      Finally, experiments at the SLC using polarized $e^-$ beams
      could measure the ratio of the $e^-_L$ and  $e^-_R$  couplings
      directly from the relative rates of the production of hadronic $Z$ events.    The
      value of $A_e$ inferred from this technique was
      \beq
         A_e = 0.1516 \pm 0.0021  \ . 
         \eeq{AeSLD}

\begin{figure}
\begin{center}
\includegraphics[width=0.60\hsize]{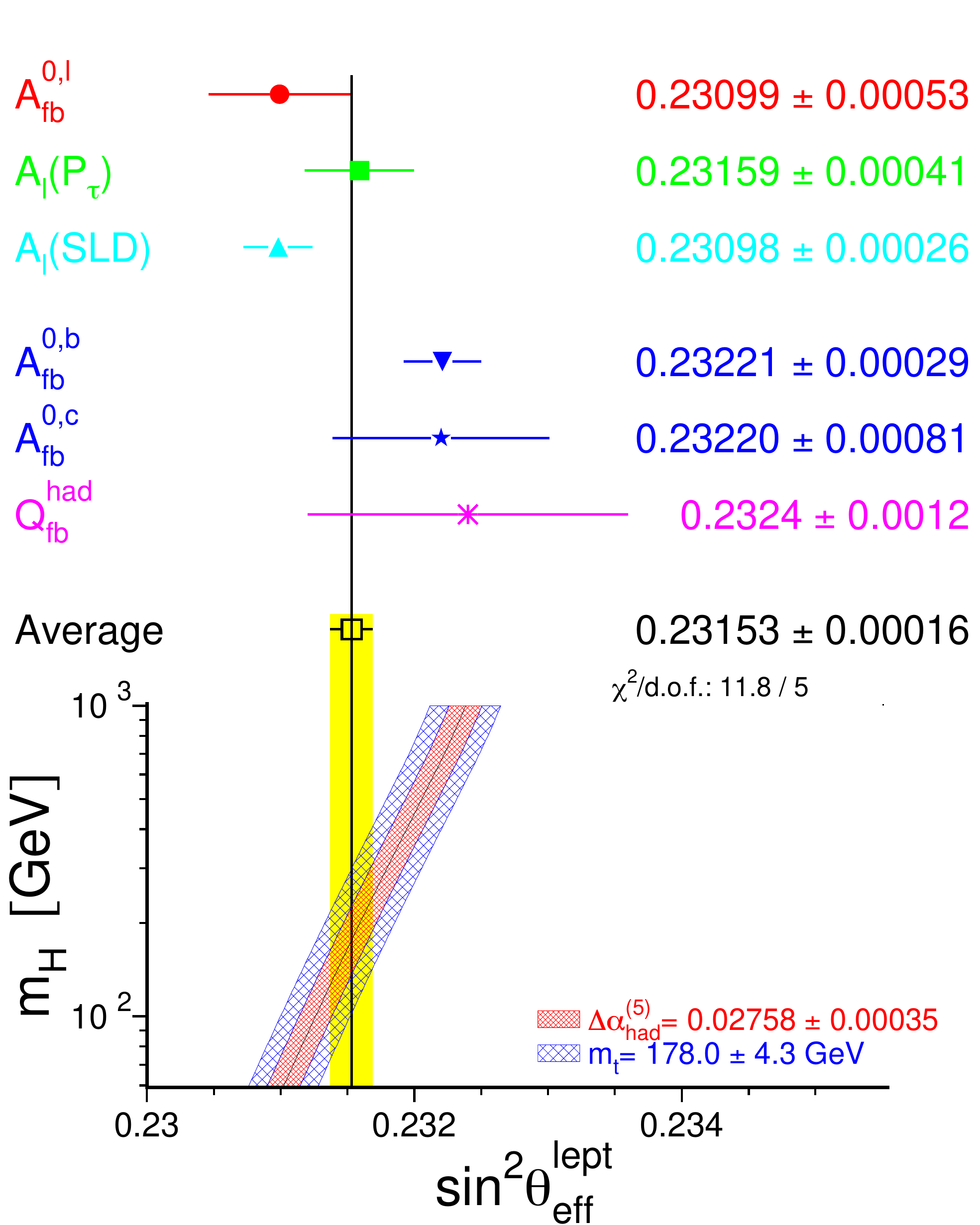}
\end{center}
\caption{Comparison of the measurements of $A_\ell$ from a variety of
  precision electroweak measurements \cite{LEPEWZ}.   The measurements
  are expressed in terms of the effective electroweak mixing angle
  $\sin^2\theta_{*}$. }
\label{fig:Aell}
\end{figure}

\begin{figure}
\begin{center}
\includegraphics[width=0.95\hsize]{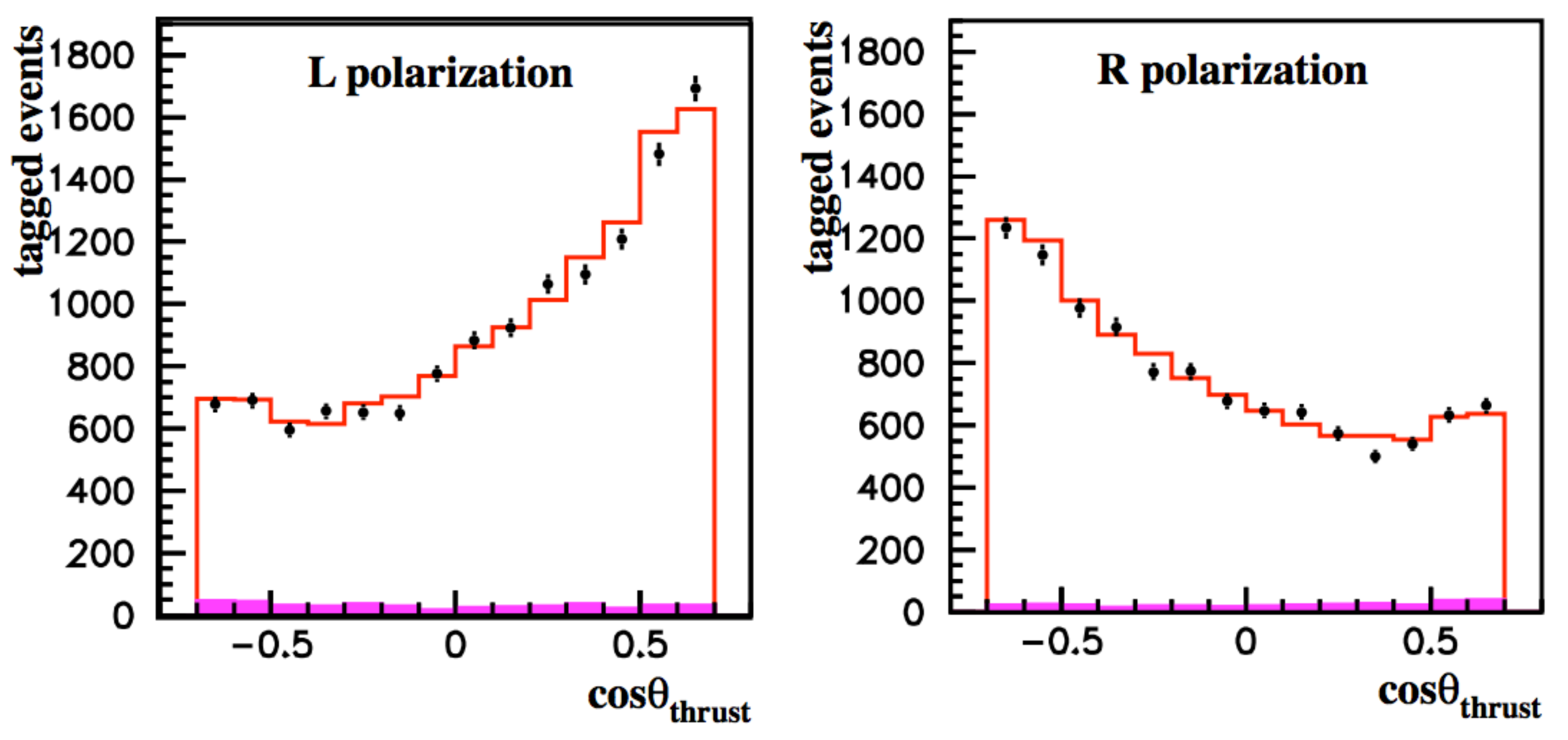}
\end{center}
\caption{Measurement of the angular distribution of $Z\to b\bar b$
  events by the SLD experiment using polarized electron beams~\cite{SLDb}.}
\label{fig:Ab}
\end{figure}

   The overall consistency of the various $A_\ell$ measurements is
   shown in Fig.~\ref{fig:Aell}~\cite{LEPEWZ}.   The various
   measurements are expressed as values of $\sin^2\theta_*$, defined
   by  \leqn{twomoresstw}.   Much ink has been spilled over the
   3~$\sigma$
   difference in the measurements from $A_\ell(SLD)$ and $A_{FB}^b$.
   This difference does not influence the overall
   confirmation of the Standard Model.   Still, it would be good to
   measure the $A_\ell$ even  more accurately. That can be done at
   next-generation $\ee$ colliders~\cite{FCCee,ILC}.

   In contrast to $A_\ell$, the value of $A_d$ or $A_b$ in the
   Standard Model is expected to be almost maximal.   This prediction
   can be tested using polarized beams.  Maximality implies that the
   distribution of $b$ quarks in $e^-e^+ \to b\bar b$ should be
   strongly forward-peaked for $e^-_L$ beams and strongly
   backward-peaked for $e^-_R$ beams.    The results of measurements
   at the SLC with polarized beams confirm this prediction, as shown
   in Fig.~\ref{fig:Ab}~\cite{SLDb}.

\begin{figure}
\begin{center}
\includegraphics[width=0.80\hsize]{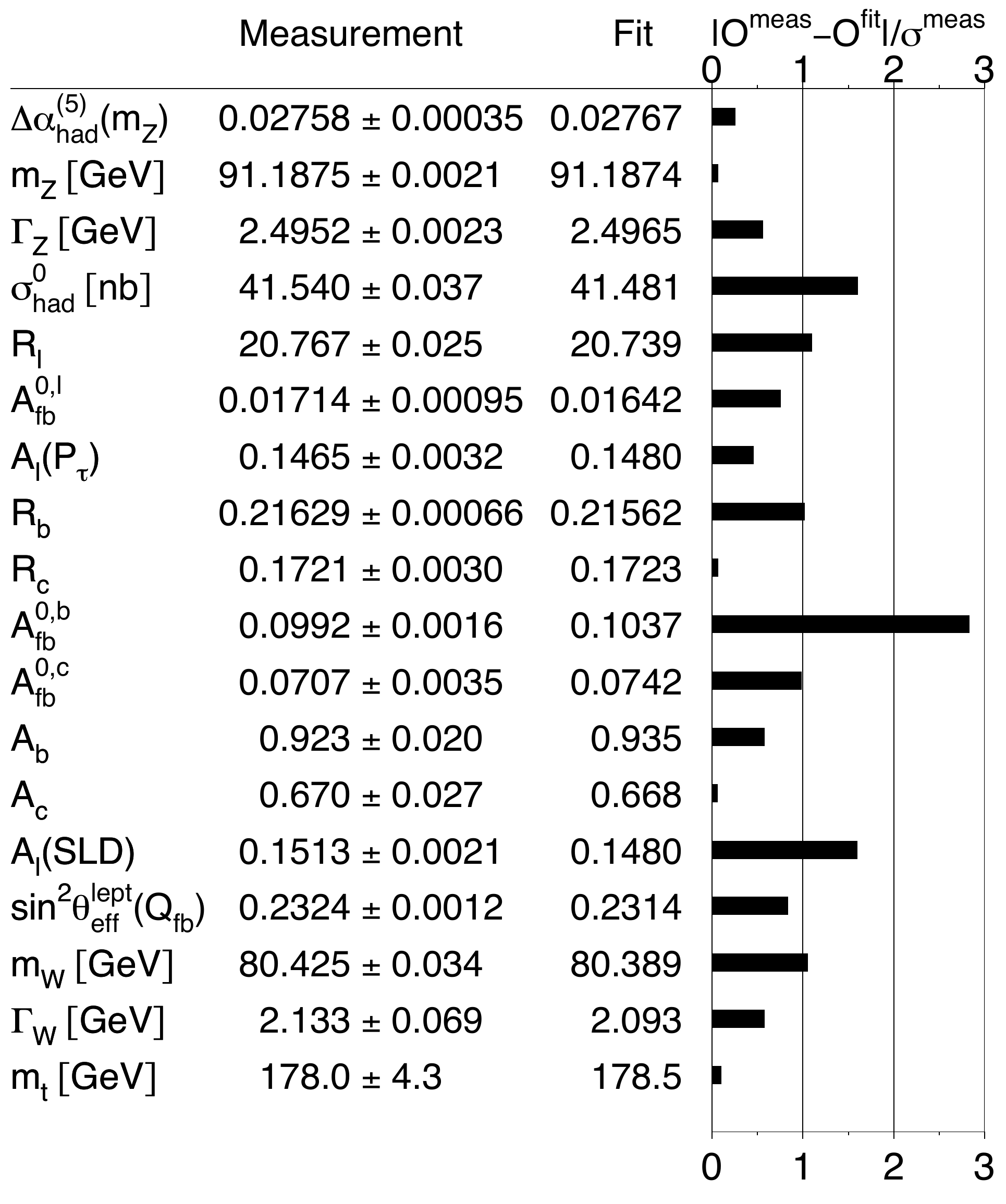}
\end{center}
\caption{Compilation of precision electroweak measurements, and
  comparison to the predictions of the Standard Model using the
  best-fit parameters~\cite{LEPEWZ}.}
\label{fig:EWfit}
\end{figure}

   The overall compatibility of the measurements of electroweak observables with the
   values predicted by the best-fit Standard Model is shown in
   Fig.~\ref{fig:EWfit}~\cite{LEPEWZ}.    The bars on the right show
   the deviations of each measurement from the Standard Model
   expectation, in $\sigma$.

   \section{Prediction of $m_t$}
   In Section~\ref{sec:sstw}, I noted that the largest radiative
   correction to electroweak predictions comes in the renormalization
   of the value of $\alpha$ from 1/137 at $Q^2 = 0$ to 1/129 at
   off-shell momenta of order $m_Z^2$.    There is another source of
   relatively large corrections:  Since the top quark is a heavy quark
   with mass much greater than the $W$ mass,  loops containing the top
   quark can give corrections of order
   \beq
  { \alpha_w\over 16 \pi}  {m_t^2 \over m_W^2}  \ .
   \eeq{topcorr}
The Standard Model predicts  quite significants shifts of 0.7\% in the 
$W$ boson mass and 1.3\% in the $Z$ boson width due to this effect. 

These shifts in precisely measured electroweak parameters were needed
for the success of the Standard Model fit.   Conversely, if one
assumed the validity of the 
Standard Model without additional heavy particles, the electroweak fit
put limits on the value of the top quark mass.   In the early 1990's,
as the CDF and D\O\ experiments at Fermilab raced to discover the top quark, the
Standard Model 
electroweak fit predicted an increasingly narrow range in which the
top quark should be found.   To give one example, a 1993 paper by the
ALEPH collaboration interpreted their Standard Model electroweak fit
as a measurement~\cite{ALEPHtop}
\beq
m_t =  156 \pm  {}^{22}_{25} \pm {}^{17}_{22}~\mbox{GeV} \ ,
\eeq{ALEPHtop1}
where the second error is the uncertainty associated with the unknown
value of the Higgs boson mass~\cite{ALEPHtop}.  A 1994  update of this analysis by
Martinez
gave~\cite{Martinez}
\beq
m_t =  156  {}^{+22}_{-25} {}^{+17}_{-22}~\mbox{GeV} \ ,
\eeq{ALEPHtop2}
an estimate quite comparable to that given by the CDF Collaboration in
1994 at the ``evidence for'' stage of the direct search for the top
quark~\cite{CDFtop},
\beq
m_t =  156 \pm  10 {}^{+13}_{-12}~\mbox{GeV} \ ,
\eeq{CDFtop1}

\begin{figure}
\begin{center}
\includegraphics[width=0.60\hsize]{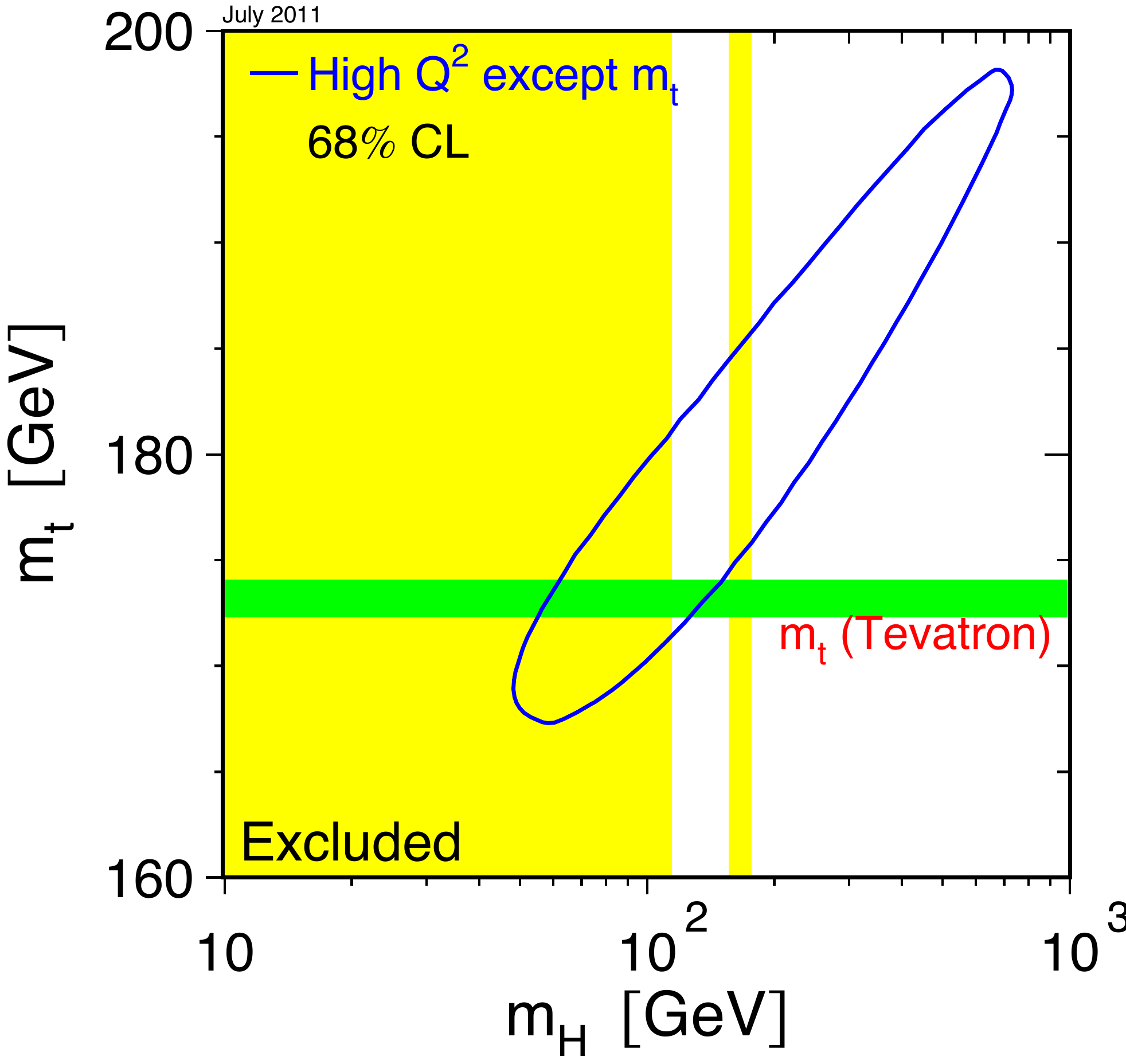}
\end{center}
\caption{68\% confidence region of the $m_h$-$m_t$ plane allowed by
  precision electroweak measurements, as determined by the LEP
  Electroweak Working Group in 2011~\cite{LEWWG2011}.
  The measurement of $m_t$ and the constraints on $m_h$ are from the
  Tevatron experiments CDF and D\O.}
\label{fig:mtmh}
\end{figure}

It is much more difficult to obtain strong constraints on the mass of
the Higgs boson.   Loop diagrams containing the Higgs boson depend on
the Higgs boson mass only as
  \beq
  { \alpha_w\over 16 \pi} \log {m_h^2 \over m_W^2}  \ .
   \eeq{Higgscorr}
This is  a much weaker dependence on the unknown mass and also gives
shifts about 5 times smaller for the actual mass values.   However, it
became clear in the late 1990's that values of the Higgs boson mass
above about 200~GeV would give results inconsistent with the observed
values of the 
electroweak observables.   Figure~\ref{fig:mtmh}, produced by the LEP
Electroweak Working Group in 2011, shows the contemporary 68\% CL contour in the
plane of $(m_h, m_t)$~\cite{LEWWG2011}.    The constraint from the
known value of the top quark mass clearly indicated a value of the
Higgs boson mass below 200~GeV.  This constraint did assume the
validity of the Standard Model.   Theories with physics beyond the
Standard Model could allow a heavy Higgs boson, but only in specific
scenarios with well-defined and observable
consequences~\cite{WellsP}.   We now know that these scenarios did not
play out and that the prediction based on the Standard Model was correct.

\section{$\ee\to W^+W^-$}

I should not end this discussion of the successes the Standard Model
in describing electroweak interactions without noting one other
important LEP measurement, that of the cross section for the process
$\ee\to W^+W^-$.

\begin{figure}
\begin{center}
\includegraphics[width=0.50\hsize]{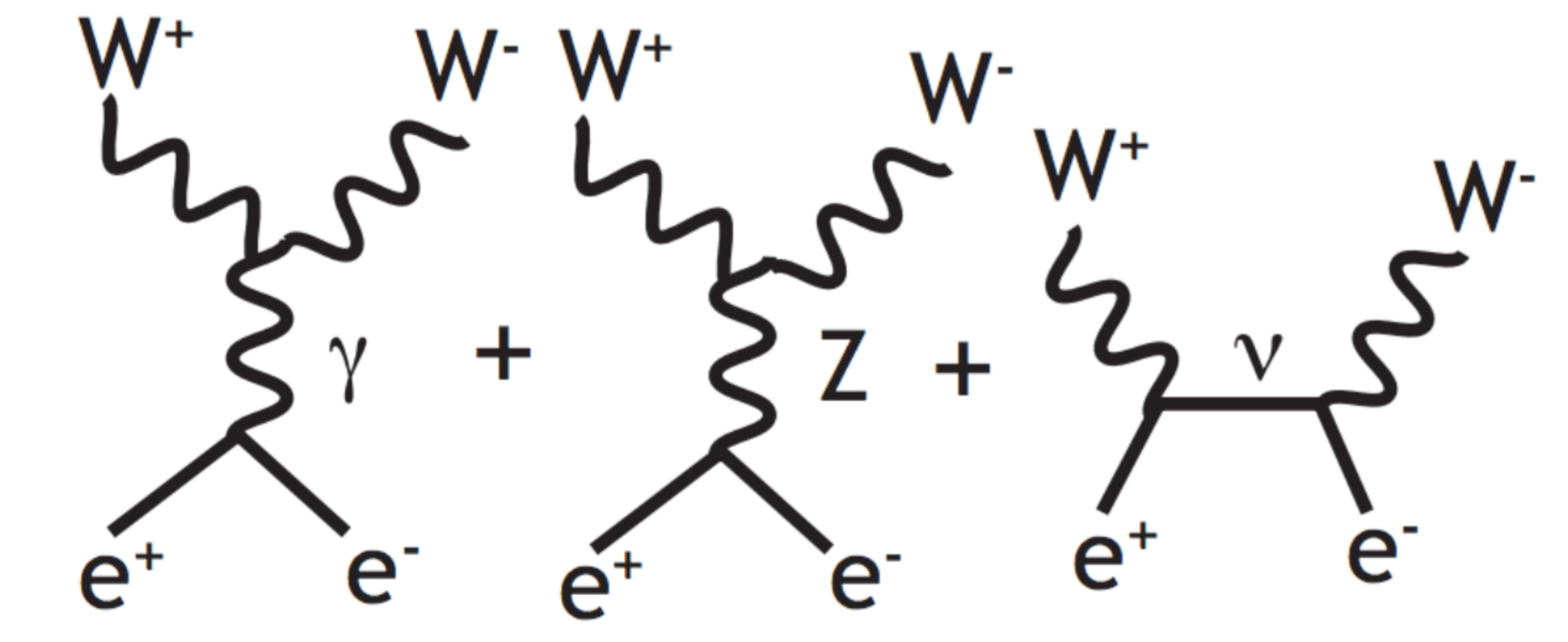}
\end{center}
\caption{Diagrams contributing to the amplitude for $\ee\to W^+W^-$ at
  the tree level.}
\label{fig:WWdiagrams}
\end{figure}

Thinking naively, the amplitude for the production of two longitudinally
polarized $W$ bosons should be approximated by the cross section for
the production of charged scalars, times the product of the $W$
polarization vectors,  giving
\beq
    {d\sigma\over d\cos\theta}(\ee\to W^+_0W^-_0)\sim 
 {d\sigma\over d\cos\theta}(\ee\to w^+_0w^-_0) \cdot
\bigl| \epsilon^{*\mu}(W^+)
\epsilon^{*}_{\mu}(W^-)\bigr|^2
\eeq{naive}
However, the product of these polarization vectors becomes very large
at high
energy,
\beq
\epsilon^{*\mu}(W^+)
\epsilon^{*}_{\mu}(W^-)  \to    {s\over 2 m_W^2} \ ,
\eeq{eeasymp}
leading to a prediction for the S-wave cross section that violates
unitarity.  In the Standard Model, the amplitude for $\ee\to W^+W^-$
is given at the tree level by sum of the three diagrams shown in
Fig.~\ref{fig:WWdiagrams}.   Thus, there is the possibility that the
unitarity-violating terms could cancel among these diagrams.  At first
sight, these seems unlikely.   But, in fact,
this cancellation is guaranteed by $SU(2)\times
U(1)$ gauge invariance and its consequence, the Goldstone Boson
Equivalence Theorem~\cite{GB1,GB2,GB3}.    The tree-level analysis was
done in the mid-1970's by Flambaum, Khriplovich, and Sushkov and by
Alles, Boyer, and Buras~\cite{Flambaum,Buras}, and their results
displayed this cancellation explicitly.

\begin{figure}
\begin{center}
\includegraphics[width=0.26\hsize]{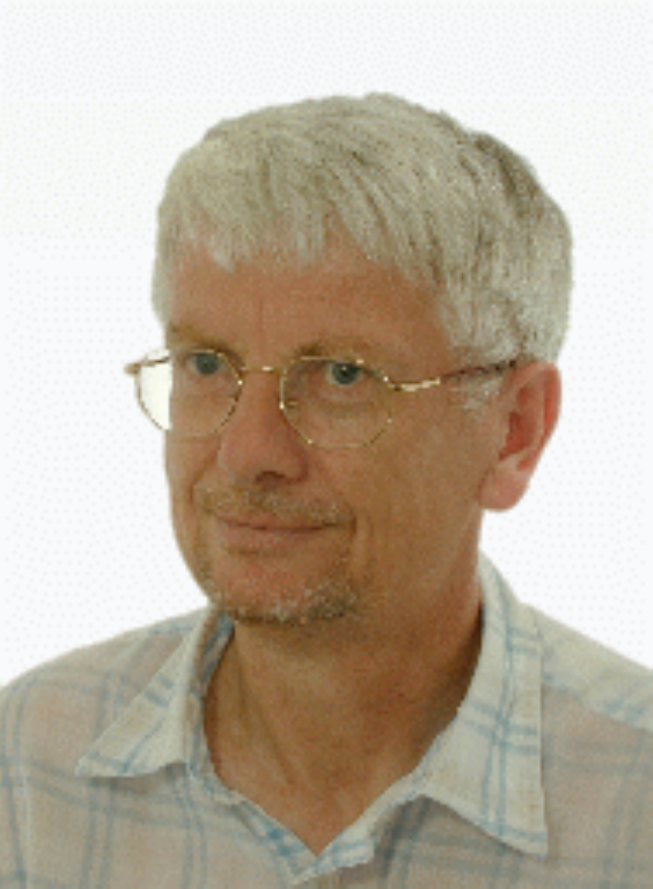} \qquad
\includegraphics[width=0.28\hsize]{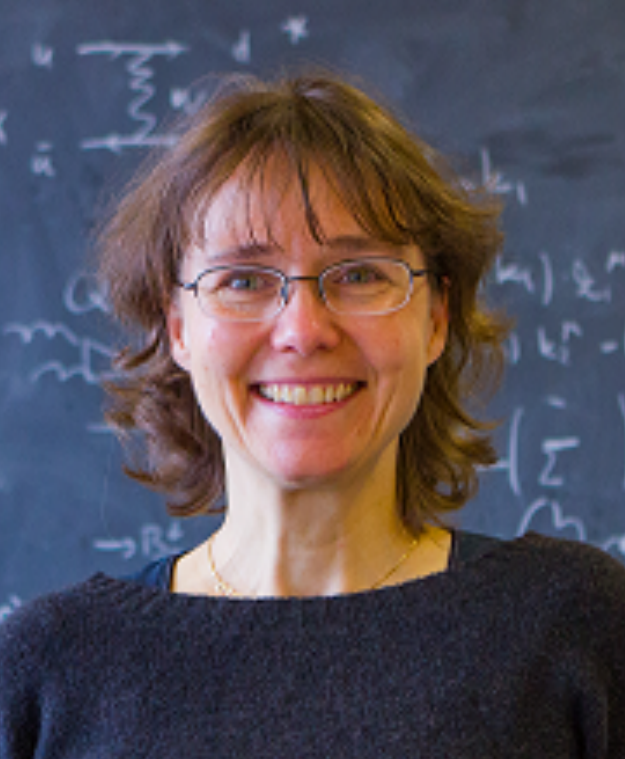} 
\end{center}
\caption{Authors of precision event generators for $\ee\to 4$~fermion
  processes: (left) Stanislaw Jadach, the senior author of YFSWW,
  (right) Doreen Wackeroth, the junior author of RACOONWW.}
\label{fig:Doreen}
\end{figure}

\begin{figure}
\begin{center}
\includegraphics[width=0.90\hsize]{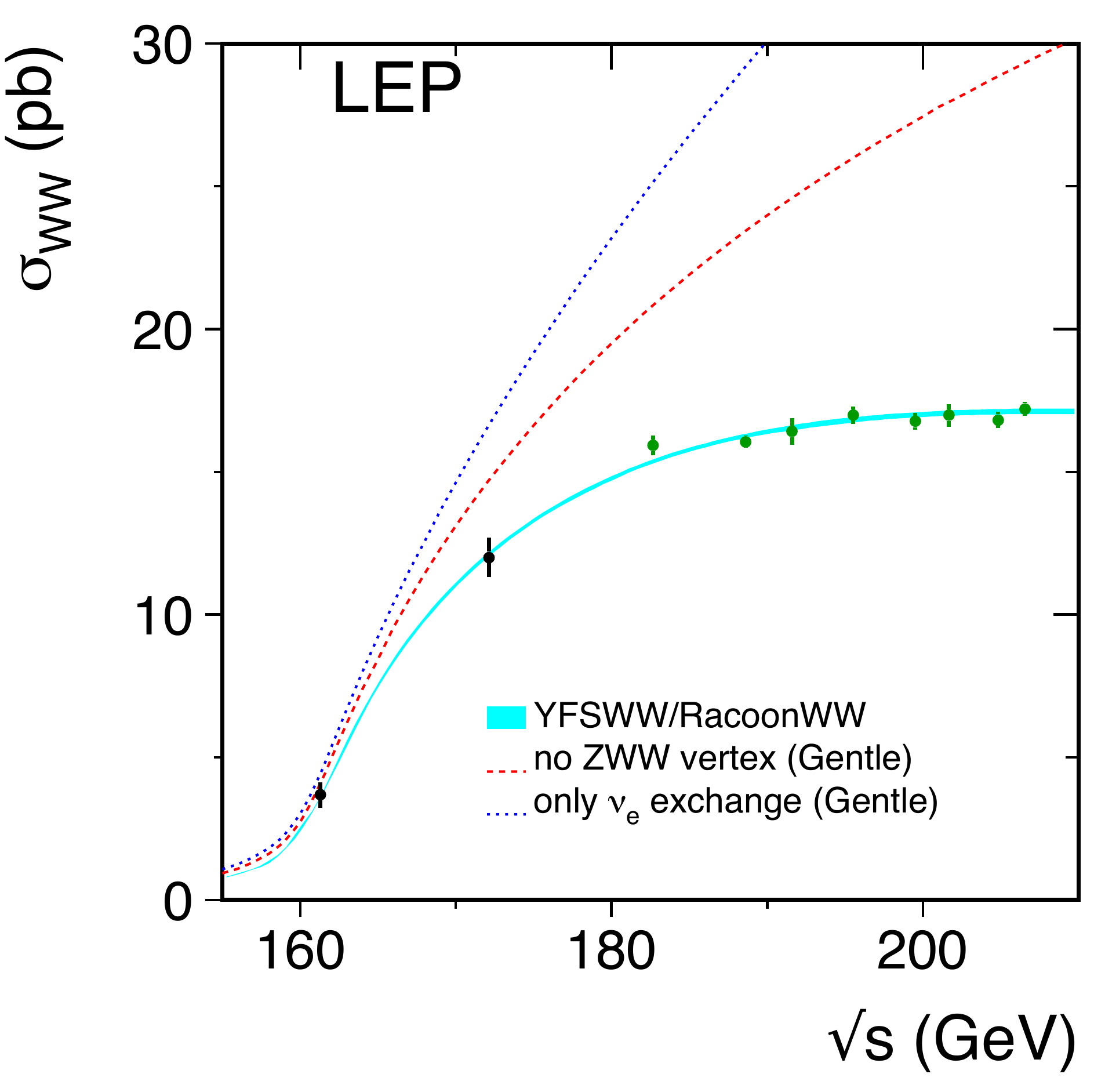}
\end{center}
\caption{Measurements of the total cross section for $\ee\to W^+W^-$
  by the four LEP experiments, compared to theoretical predictions~\cite{LEPEWWW}.}
\label{fig:WW}
\end{figure}

Providing  a theoretical prediction for the precision measurement of
this cross section at LEP 2 made it necessary to solve  additional
problems.   To treat $W$ bosons realistically, it is necessary to
allow them to go off the mass shell, while retaining a sufficient level
of gauge-invariance to allow the gauge cancellations to go through.
Processes with
off-shell $W$ bosons are in principle indistinguishable from general
$\ee\to$~4~fermion processes, and so these must also be modelled
correctly in the event generator.   These issues were addressed in the
codes YFSWW, by Jadach, Placzek, Skrypek, Ward, and
Was~\cite{KORALW1,KORALW2},  and
RACOONWW,
by Denner, Dittmaier, Roth, and
Wackeroth~\cite{RACOON1,RACOON2,RACOON3}.  

The results of the four LEP 2 experiments, compared to the predictions
of the event generators, are shown in Fig.~\ref{fig:WW}.  For
illustration, the predictions of tree-level calculations that omit the
$ZWW$ diagram and both of the $s$-channel diagrams in
Fig.~\ref{fig:WWdiagrams} are also shown.    The cancellation required
by gauge invariance is manifest.   This cancellation takes place only
if the triple gauge boson vertices are of the form required by
Yang-Mills theory within an accuray of a few percent.

\section{Conclusions}

The LEP/SLC program of precision electroweak measurement required an
unprecedented theoretical effort to provide high-precision predictions
of the properties of the electroweak bosons.   The comparison of these
theoretical and experimental efforts was a triumph that leaves no
doubt that $SU(2)\times U(1)$ gauge invariance is actually a property
of nature.   Whatever lies beyond, we can now take the validity of
electroweak gauge invariance as a foundation to rely on as we move forward.

\Acknowledgements

I am grateful to  Bryan Lynn, Harsh Mathur, Glenn Starkman, Kellen
McGee,
and their team at
Case Western Reserve for inviting me to this symposium, and for making
it so memorable.   I am grateful to Bryan Lynn for his patient
introduction to precision electroweak theory  and to many illuminating
discussions over the years.   I thank Alain Blondel, Lance Dixon,
Morris Swartz, and many other colleagues at SLAC and elsewhere for
assisting my education in this subject.  This work was supported by
the
U.S. Department 
of Energy under contract DE--AC02--76SF00515.


\begin{thebibliography}{99}
\bibitem{Higgs}
  In processes that involve the Higgs boson, a fourth parameter
  enters, the Higgs self-coupling $\lambda$.    This can be fixed by
  an on-shell renormalization that sets the Higgs boson mass to its
  observed value.
\end{thebibliography}
\end{document}